\documentclass[grl]{agu2001}
\usepackage{agu-ps}
\usepackage[dvips,final]{graphicx}
\usepackage{amssymb}
\usepackage{epsfig}
\ifgalley
\setlastpagenum{4}
\fi
\newcommand{\be}{\begin{equation}}
\newcommand{\ee}{\end{equation}}

\newcommand{\eqref}[1]{eq.\,(\ref{#1})}

\newcommand{\figwidth}{\hsize}

\authorrunninghead{COSTA} 
\titlerunninghead{CRYSTAL-CONTENT VISCOSITY DEPENDENCE}

\journalid{}
\articleid{}{}
\cpright{agu}{2005}
\revised{}
\accepted{}
\published{}

\authoraddr{A. Costa,
Department of Earth Sciences, University of Bristol, Wills Memorial
Building, Queen's Road, Bristol BS8 1RJ - UK. 
(e-mail: \hbox{a.costa@}bris.ac.uk)}
\slugcomment{Submitted in the {\it Geophysical Research Letters}, 2005.}

\begin{document}
%\title{Solid-fraction viscosity dependence in melts with high
%  crystal contents}   
\title{Viscosity of high crystal content melts: dependence
  on solid fraction} 
\author{Antonio Costa}
\affil{Centre for Environmental and Geophysical Flows, Department of Earth
Sciences, University of Bristol}

\begin{abstract}
The rheological properties of suspensions containing high solid
fractions are investigated. Attention is focused on viscosity of 
silicate and magmatic melt systems. A general empirical equation which
describes the relative viscosity of suspensions as a function of
suspended solid fraction is proposed. In the limit of very dilute
solid concentrations it reduces to the Einstein equation. The proposed
relationship is satisfactorily applied to reproduce available
experimental data relative to silicate melts. Moreover, the
extrapolation of the model to very high concentrations is compared
with experimental observations on partially-melted granite.      
\end{abstract}
\begin{article}
\section{Introduction}
The rheology of suspensions with high solid fraction is of
interest in many fields, from industry \citep{liu2000} to magma
transport processes \citep{lejric95}. Regarding magma rheology at high
solid fractions, there is little reliable experimental data about
crystal-content viscosity dependence, even though magma viscosity is a   
fundamental property governing mass transport in volcanological
processes. This is mainly due to the difficulty in performing
experiments on magma containing crystals, as well as to the difficulty
in interpreting the results because of the numerous factors that
influence the rheology of partially crystallised magma
\citep{pinste92,lejric95,saaman2001}. However, since magma
viscosity controls magma transport, when modelling many volcanic
processes (e.g. dome growth) there is a need to estimate
crystal-content viscosity dependence even in the limit of very high
crystal content where no experimental observations are available
\citep[see e.g.][]{melspa99,melspa2002}. \\ 
In the first part of this paper I shortly review the most common 
parameterisations proposed for the rheology of concentrated
suspensions, then I propose a new equation describing the relative
viscosity as function of the solid content. Such an equation 
is able to reproduce the available experimental data, reducing
to the well established relationships for dilute suspensions. Finally,
I apply the proposed parameterisation to fit data on silicate
melts \citep{lejric95} and to describe the trend of viscosity of
partially-melted granite \citep{molpat79}.    
\section{Viscosity of very concentrated suspensions}
One century ago \citet{ein06} predicted that in the case of a
very diluted suspension of solid spherical non-interacting particles,
the relative viscosity $\eta$ (defined as the ratio of suspension
viscosity on that of the suspending medium) is  
\be
\eta= (1+B\phi)
\label{einstein}
\ee
where $\phi$ is the volume fraction of particles (defined as the 
ratio of volume occupied by particles on total volume of suspension){\bf,}
and $B$ is the Einstein coefficient with a theoretical value $B=2.5$
\citep[from experiments for $\phi\rightarrow 0$, it ranges from 1.5 to 
5,][]{jefacr76}. 
Considering eq.~(\ref{einstein}) as a starting point, many
rheologists have attempted to find a formula for the effective
viscosity $\eta$ in terms of $\phi$, i.e. $\eta=\eta(\phi)$. For
spherical particles having different sizes, \citet{ros52} and
independently \citet{bri52} showed that (\ref{einstein}) implies a
relative viscosity given by:  
\be
\eta= \left(1-\phi\right)^{-2.5}
\label{roscoe}
\ee
In the derivation of eq.~(\ref{einstein}) particle interactions
are completely neglected. In order to extend the validity of
(\ref{roscoe}) to suspensions containing a high concentration of
spheres, \citet{ros52} introduced the concept of critical fraction 
$\phi_m$ at which incompressible solid would prevent any flow:
\be
\eta= \left(1-\frac{\phi}{\phi_m}\right)^{-2.5}
\label{roscoe2}
\ee
For spheres of equal size, the relative maximum packing density is
$\phi_m\simeq 0.74$, for a random loose packing is around 0.60,
whereas for spheres of very diverse sizes $\phi_m$ cannot be greater
than $\simeq 0.93$ \citep[$\phi_m\simeq 0.87$ for a polydisperse
random sphere packing,][]{kantor2002}. Eq.~(\ref{roscoe2}) is an
attempt to describe $\eta$ in terms of  $\phi$ starting from first
principles, but as it has been shown \citep{jefacr76}, no simple
single functions (i.e. $\eta=\eta(\phi)$) can exist because suspended 
particles can be subjected to thermal, electrical and hydrodynamic
interactions and moreover particle shapes and particle size
distribution play an important role that cannot be easily incorporated
into a simple form. At present, there is no rigorous theory
\citep{herpie80}. For these reasons any proposed function that can fit
experimental data must contain adjustable parameters. For instance, a
semi-empirical generalisation of (\ref{roscoe2}) due to
\citet{kridou59} is:   
\be
\eta= \left(1-\frac{\phi}{\phi_m}\right)^{-B\phi_m}
\label{roscoe3}
\ee
that in the limit $\phi\rightarrow 0$ reduces to $1+B\phi$. In this
way $B$ can be interpreted as an Einstein coefficient. Actually,
$\phi$ should be viewed generally as an effective volume fraction
\citep{que98}.
For concentrated suspensions, no satisfactory theory comparable with
that for diluted concentrations exists \citep{jefacr76}, but a variety
of empirical equations have been suggested for description of
$\phi-\eta$ relationship 
\citep[see e.g.][]{gaynel69,dabyuc86,liu2000}. 
All the proposed parameterisations for concentrated suspensions assume
that there is a critical solid fraction $\phi_m$ at which the
viscosity tends to be infinite. From a physical point of view it is
instead natural to assume that when the solid fraction reaches $\phi_m$
a rheological transition occurs from a regime where the rheology is
basically determined by the liquid phase to a regime where the effect
of crystals is predominant and the viscosity values are much higher
\citep{lejric95,molpat79}. In the latter regime the rheology is more
complicated than in the first one, the suspension becomes markedly 
non-Newtonian exhibiting a yield stress and the liquid-solid system
can behave like a fluid or a solid depending on the stress applied on
it. Only an effective viscosity value can be inferred for these
systems \citep{molpat79}. Experimental results obtained by
\citet{lejric95} for crystal-bearing melts clearly suggest this
rheological transition. \\  
Based on this idea, in order to reproduce this transition between
the two different rheological regimes, \citet{melspa99,melspa2002}
utilised the following parameterisation:
\be
\eta=\theta_0\exp{\left\{\mbox{arctang}[\omega(\phi-\phi_*)]+
\frac{\pi}{2}\right\}} 
\label{melspa}
\ee
where $\theta_0$, $\omega$ and $\phi_*$ are three adjustable parameters
that must be chosen by best fitting. We must note that in the
limit of small $\phi$, (\ref{melspa}) does not recover
eq.~(\ref{einstein}).  \\ 
Our purpose in the next section is to find a more general 
$\phi-\eta$  parameterisation that (i) describes the aforementioned
rheological transitions, (ii) is able to reproduce the few available
data of viscosity for high crystal content silicate melts
\citep{lejric95}, (iii) contains adjustable parameters easy to
interpret from a physical point of view, and (iv) in the limit of
small $\phi$ reduces to the equations valid for dilute systems.
\section{Proposed parameterisation} 
\begin{figure}
\centerline{\includegraphics[angle=0,height=0.6\figwidth]{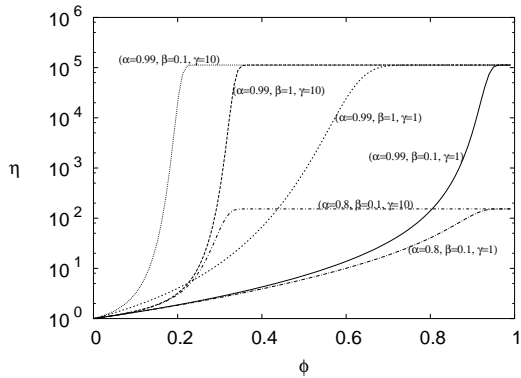}}
\caption{Graphic of the relationship~(\ref{costa}) for different
  parameter values. Triplets near the curves indicate the
  corresponding values of ($\alpha$, $\beta$, $\gamma$).}
\label{graph}
\end{figure}

Considering  eq.~(\ref{roscoe2}) as the starting point, we assume
a relationship $\phi-\eta$ of the following form:
\be
\eta=\left[1-F(\phi,\alpha,\beta,\gamma)\right]^{-B/\alpha}
\label{step1}
\ee
where $B$ is Einstein's coefficient (theoretically $B=2.5$),
$\alpha,\beta,\gamma$ are three adjustable parameters (they should
allow us to reproduce a linear, an intermediate and an asymptotic
behaviour) and $F$ is a general function which has a rapid asymptotic
behaviour for large $\phi$  whereas is linear as 
$\phi\rightarrow 0$. As a general function we choose $F$ proportional
to the error function, i.e. the integral of the Gaussian distribution:
$\mbox{erf}(x)\equiv
\frac{2}{\sqrt{\pi}}\int_0^x\exp{\left(-t^2\right)}dt$ 
limited at $x \ge 0$. 
This function has the appropriate features mentioned above and the
Taylor expansion near $x=0$ gives:   
\be
\mbox{erf}(x)=\frac{2}{\sqrt{\pi}}x-\frac{2}{3\sqrt{\pi}}x^{3}+O(x^4) 
\label{erf2}
\ee 
We assume:
\be
\eta=\left[1- \alpha\,\mbox{erf}(\phi,\beta,\gamma)\right]^{-B/\alpha}
\label{step2}
\ee
where the argument of the error function must contain a term linear in
$\phi$ and a non-linear term which permits a rapid saturation for
large $\phi$ and it is able to incorporate the non-linear corrections
in the intermediate range. The final form we adopt is the following:
\be
\eta(\phi)=\left\{1- \alpha\,\mbox{erf}
\left(\frac{\sqrt{\pi}}{2}\phi\left[1+\frac{\beta}{(1-\phi)^\gamma}\right]
\right)\right\}^{-B/\alpha}
\label{costa}
\ee
where $0<\alpha<1$. In this way, if it needs, the relative viscosity
can assume all the value from 1 to infinity (see Figure~\ref{graph}
where relationship~(\ref{costa}) is shown for different parameter
values).   
When the argument of the function erf($x$) is $<1$ (i.e. as we can
neglect terms of the third order in the Taylor expansion),
relationship~(\ref{costa}) tends to: 
\be
\eta(\phi)=\left\{1- \alpha\,\phi\left[1+\frac{\beta}{(1-\phi)^\gamma}\right]
\right\}^{-B/\alpha}
\label{costa1}
\ee
The critical fraction $\phi_c$ at which a sharp rheological change
occurs can be estimated considering that value of $\phi$ at which the
argument of the function erf($x$) is close to the unit:
\be
\phi\left[1+\frac{\beta}{(1-\phi)^\gamma}\right]
\approx  \frac{2}{\sqrt{\pi}}
\label{phic}
\ee   
The numerical solution of (\ref{phic}) furnish an estimate of
$\phi_c$. For $\phi\lesssim \phi_c$ we can recover a relationship
similar to (\ref{roscoe3}):
$\eta(\phi)\approx\left(1-\alpha\,{\phi}/{\phi_c}\right)^{-B/\alpha}$.
Finally, for very small $\phi$ (i.e. as we can neglect terms of the
second order in the Taylor expansion), eq.~(\ref{costa}) becomes:  
\be
\eta(\phi)\simeq\left(1+B_*\phi\right)
\label{costa2}
\ee
where $B_*=(1+\beta)B$ is an Einstein coefficient and $\beta$ can be
viewed as a correction to the theoretical value of $B$ due to the
non-linear term in the argument of $F$. 
\par
A physical interpretation of the empirical parameters that appear in
eq.~(\ref{costa}) can be obtained from the relationships
written above and from Figure~\ref{graph}. Concerning $\alpha$,
directly  from (\ref{costa}) we can see that it gives the maximum
relative viscosity in the limit of high fraction solid: 
$\eta_{max}\equiv(1-\alpha)^{-B/\alpha}$.  
$\beta$ and $\gamma$ parameters indicate the importance of non-linear
effects in the argument of the erf$(x)$ function.
Figure~\ref{graph} shows that $\beta$ mainly controls the value
of the critical fraction $\phi_c$ (small $\beta$ values correspond to
smaller $\phi_c$). From (\ref{costa2}), it is possible to impose for
$\beta$ an upper limit $\lesssim 2$ in order to maintain $B_*$ within
the typical range measured in experiments. Finally, from
Figure~\ref{graph} we can see that $\gamma$ (here assumed to be 
greater than unity) determines the curve slope between the two regimes
(a measure of the ``rapidity'' to reach the asymptotic viscosity) and
hence the width of the transition region. Generally, we assume that
values of all three parameters $\alpha$, $\beta$ and $\gamma$ can be
shear-dependent.  
\section{Fitting of experimental data}
\begin{figure}
\centerline{\includegraphics[angle=0,width=\figwidth]{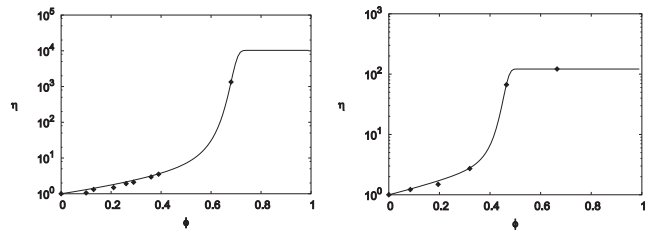}}
\caption{Measured \citep[dots]{lejric95} and predicted
  (eq.~(\ref{costa}), lines) values for viscosity-crystal
  fraction relationship in partially crystallised
  Mg$_3$Al$_2$Si$_3$O$_{12}$ (left) and Li$_2$Si$_2$O$_5$ (right). 
  Considering a relative error of about 15\%, the best fitted values
  $\alpha=0.9725;\,0.7735 $ (i.e. $\eta_{max}=1.03\times
  10^4;\,1.21\times 10^2$), $\beta=0.0096;\, 0.0014$, $\gamma=4.43;\,
  11.81$ give $\chi^2=0.8;\, 0.25$  for Mg$_3$Al$_2$Si$_3$O$_{12}$ and 
  Li$_2$Si$_2$O$_5$  respectively . The correlation
  coefficient is practically equal to 1 in both cases.}  
\label{fit}
\end{figure}

For what concerns viscosity of partially crystallised silicate melts
as a function of crystal volume fraction, only very few data are
available, the most significant being by \citet{lejric95}. In
their experimental study, \citet{lejric95} measured the viscosity of
partially crystallised Mg$_3$Al$_2$Si$_3$O$_{12}$ and
Li$_2$Si$_2$O$_5$. Crystals were well-rounded in the first case
while the second kind of melt contained small ellipsoidal inclusions,
but the observed rheological behaviour was qualitatively similar
in both cases. Their results suggest that the most important parameter
in determining melt viscosity is basically the crystal fraction,
although other factors can influence the rheological behaviour.\\ 
As it is shown in Figure~\ref{fit}, the agreement between experimental
data of \citet{lejric95} and  results given  by best fitting
eq.~(\ref{costa}) is excellent. Using the least square minimisation
procedure, in both cases we found  very low $\chi^2$ values 
%($\chi^2<1$ considering a total error of 15\%) 
and a correlation coefficient practically equal to the unit which
indicate the elevated goodness of the fits. Besides these results, the
numerical solution of eq.~(\ref{phic}) for the estimation of the
critical crystal fraction, with the parameter values reported in
Figure~\ref{fit} furnishes $\phi_c\approx 0.65$ and $\phi_c\approx
0.45$ for Mg$_3$Al$_2$Si$_3$O$_{12}$ and Li$_2$Si$_2$O$_5$ melts
respectively. In the case of Li$_2$Si$_2$O$_5$, fitting results show
that nonlinear terms are more important with respect to
Mg$_3$Al$_2$Si$_3$O$_{12}$ while the lower $\alpha$ could reflect the
higher zero crystal fraction melt viscosity. The different behaviour
of Li$_2$Si$_2$O$_5$ melts could be due to the presence of ellipsoidal
particles formed during crystallisation (in the case of
Mg$_3$Al$_2$Si$_3$O$_{12}$ particles were well-rounded
spherulites). Finally, in the limit of $\phi\ll 1${\bf,} for both
cases the coefficient $B_*$ in (\ref{costa2}) remains practically
equal to 2.5.\\ 
Notwithstanding the model showed a good performance, we must
stress its intrinsic limitations specially in describing natural
systems containing crystals of different shapes and sizes. With
increasing crystal fraction, the stress-strain rate relationship
becomes increasingly non-linear because of interactions between
particles that create a non-Newtonian behaviour that can be
empirically described in terms of an effective viscosity and of a
yield strength. Moreover there are no reliable experimental results
for $\phi$ greater than 0.7, hence values predicted by (\ref{costa})
in the high concentration limit must be seen as extrapolations. For
these reasons more experimental data of different melt-crystal systems 
for more thorough future investigations and calibrations are
necessary.
\section{Application to melted rocks}
\begin{figure}[htb]
\centerline{\includegraphics[angle=0,height=0.6\figwidth]{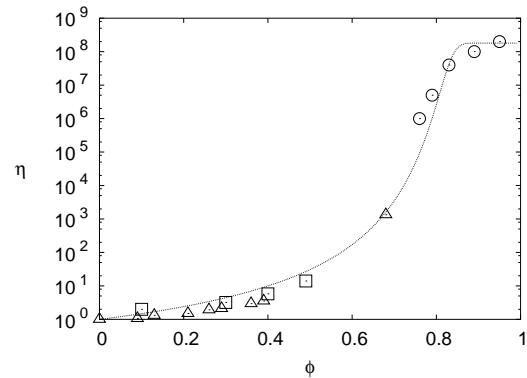}}
\caption{Effective relative viscosities deduced from data
of \citet{molpat79} at high solid fraction (circles), and from data of 
Mg$_3$Al$_2$Si$_3$O$_{12}$ by \citet{lejric95} (triangles) and other
values of \citet{tho65} at low solid fraction (squares). For the
crystal-free fluid viscosity at 800$^o$C and 300 MPa, we
considered a value of $\mu_l=10^{5}$\,Pa\,s. The correlation
coefficient is 0.94.}     
\label{total}
\end{figure}

In the case of natural systems, such as partially melted rocks,
the rheology is even more complicated because of the contemporary
influence of numerous parameters that control the viscosity
\citep[e.g.][]{mar81,pinste92}. For magma we can distinguish three 
rheological ranges: the first one for completely melted rock, an  
intermediate range and the last in the range of very low melt fraction 
\citep{molpat79}. In the first interval, and at sufficiently low
  rates of strain, melted rock rheology can be approximated to that
of an ideal Newtonian fluid with viscosity dependent on chemical
composition and temperature. In the second interval, for $\phi\lesssim
50\%$, the system can be viewed as a suspension of well separated
crystals in a viscous melt. In the low melt fraction regime,
crystal interactions have a predominant role that increases as 
the melt fraction decreases until solid-like behaviour is 
exhibited \citep{molpat79}. 
For melted rocks with high crystal percentage only a few experimental
results exist, the most significant being {those by} \citet{molpat79}
who investigated the deformation of partially-melted granite with
solid  contents from 75\% up to 95\% at 800$^o$C and 300 MPa confining 
pressure. They inferred that the critical crystal fraction
discerning granular-framework-controlled flow behaviour from
suspension-like behaviour is around 0.6-0.7. 
In a different study on rheology of lavas, from the analysis of the
overall phenocryst content \citet{mar81} observed that for crystal 
volume fractions greater than about 0.55 there should be a sharp 
rheological change that inhibits lava to erupt. With increasing
crystallinity the viscosity moderately increases but when a critical
solid content is reached, the viscosity increases so rapid that over a
short range of crystallinity the magma behaves essentially as a
solid. This should explain why lavas with more than 55\%
phenocrysts are rare.\\ 
During their experiments, \citet{molpat79} recorded some complex 
stress-strain curves having in common a generally sigmoidal
shape, but in no case did they observe a steady state
behaviour. Moreover, in the plots of maximum differential stress
versus strain-rate they note only a low strain-rate dependence (about
a tenth of an ideal Newtonian fluid). For all these reasons, it is
arduous to express their results in terms of ``effective
viscosities''. For a practical purpose, considering the results
obtained at a given strain-rate (e.g. $10^{-5}$s$^{-1}$) we define an
``effective viscosity'' as the ratio between the maximum differential
stress and the strain-rate. Keeping in mind these limitations,
in order to make a semi-quantitative comparison between the available
experimental results and eq.~(\ref{costa}), in Figure~\ref{total} we
plotted relative viscosities deduced from data of \citet{molpat79} in
the high crystal fraction regime considering as an example a
crystal-free fluid viscosity of $10^{5}$\,Pa\,s, and matched
these values with relative viscosities measured at lower solid content
relative to data of Mg$_3$Al$_2$Si$_3$O$_{12}$ by \citet{lejric95} and
some values by \citet{tho65} reported in \citet{pinste92}. Although we
cannot pretend to make an accurate fit because of the heterogeneity
and limitations of the considered data, we can observe that over
almost the entire range of solid fraction, the model is able to
reproduce the order of magnitude of all data with a good correlation
($\simeq 0.94$) and the curve shape hypothesised by
\citet{molpat79}. Moreover, considering the parameters used to plot
Figure~\ref{total} ($\alpha=0.9995$, i.e. $\eta_{max}=1.8\times 10^8$,
$\beta\simeq 0.4$ and $\gamma\simeq 1$), the numerical solution of
(\ref{phic}) gives $\phi_c \approx 0.55$, in good agreement with the
values suggested by \citet{mar81} for the sudden change of magma
rheology. 
\section{Conclusion}
A new general equation which correlates the relative viscosity of
suspensions as a function of suspended solid fraction valid for all
solid contents is proposed. The proposed relationship (i)
is able to accurately reproduce available data of viscosity for 
silicate melts with low to high crystal contents, (ii) 
describes the rheological transition between two different
regimes and (iii) contains adjustable parameters easy to interpret
from a physical point of view, and (iv) in the limit of small
solid fractions reduces to the well-known Einstein equation. 
Extrapolation to the limit of very high concentrations
satisfactorily compares with experimental observations on
partially-melted granite, giving a critical transition 
value of solid fraction around 55\%, in good agreement with
geological observations.    
\acknowledgement
This work was supported by NERC research grant reference
NE/C509958/1. The author would like to thank Prof. S. Sparks and
Dr. E. Llewellin for their interesting and useful suggestions. 
%
% *************** References ***************
%
%\bibliography{references}
%

%
\end{article}
\end{document}